\documentclass[12pt]{article}
\usepackage[usenames,dvipsnames]{color}

\usepackage{graphicx}
\usepackage{amsthm}
\usepackage{graphics}
\usepackage{amsmath,,amssymb}
\usepackage{algorithmic}
\usepackage{algorithm}
\usepackage{caption}
\usepackage{subcaption}
\usepackage{pstricks, pst-node}
\usepackage{tikz}
\usepackage[linewidth=1pt]{mdframed}
\usepackage{multirow}
\usepackage[round,colon,authoryear]{natbib}
\usepackage{pgfplots}
\usepackage{hyperref}
\usepackage{bm}
\usepackage{xr}

\pgfplotsset{compat=1.10}
\usepgfplotslibrary{fillbetween}
\usetikzlibrary{patterns}
\usetikzlibrary{shapes,arrows}
\usetikzlibrary{arrows.meta}

\newtheorem{theorem}{Theorem}
\newtheorem{proposition}{Proposition}

\newtheorem{corollary}{Corollary}

\def\EE{\mathbb{E}}
\def\PP{\mathbb{P}}
\def\RR{\mathbb{R}}

\def\Fcal{\mathcal F}

\def\Hcal{\mathcal H}

\def\bI{\mathbf I}

\def\bx{\bm x}
\def\by{\bm y}

\def\bX{\boldsymbol X}
\def\bU{\boldsymbol U}

\tikzstyle{block} = [rectangle, draw, fill=white!80!black, line width=2pt,
    text width=15em, text centered, rounded corners, minimum height=3em]
\tikzstyle{line} = [draw, -latex',line width=2pt]

\begin{document}
\title{Treatment Effects Estimation by Uniform Transformer}
\author{Ruoqi Yu$^\dagger$ and Shulei Wang$^\ast$\\ University of California, Berkeley$^\dagger$\\ University of Illinois at Urbana-Champaign$^\ast$}
\date{(\today)}

\maketitle

\footnotetext[1]{
Address for Correspondence: Department of Statistics, University of Illinois at Urbana-Champaign, 725 South Wright Street, 
Champaign, IL 61820 (Email: shuleiw@illinois.edu).}

\begin{abstract}
In observational studies, balancing covariates in different treatment groups is essential to estimate treatment effects. One of the most commonly used methods for such purposes is weighting. The performance of this class of methods usually depends on strong regularity conditions for the underlying model, which might not hold in practice. In this paper, we investigate weighting methods from a functional estimation perspective and argue that the weights needed for covariate balancing could differ from those needed for treatment effects estimation under low regularity conditions. Motivated by this observation, we introduce a new framework of weighting that directly targets the treatment effects estimation. Unlike existing methods, the resulting estimator for a treatment effect under this new framework is a simple kernel-based $U$-statistic after applying a data-driven transformation to the observed covariates. We characterize the theoretical properties of the new estimators of treatment effects under a nonparametric setting and show that they are able to work robustly under low regularity conditions. The new framework is also applied to several numerical examples to demonstrate its practical merits.
\end{abstract}



\newpage
\section{Introduction}
\label{sc:intro}


In order to infer causal relations in an observational study, a major difficulty is to reduce the bias brought by the confounding covariates related to both the treatment assignment and the outcome of interest \citep{imbens2015causal}. This task can be accomplished by balancing the empirical distributions of observed confounders in different treatment groups. One common strategy to adjust imbalances of confounders is weighting \citep{rosenbaum1987model,robins1994estimation,robins2000marginal,hirano2001estimation,hirano2003efficient}, which seeks a weight for each sample so that covariates distributions are more similar between the weighted groups. 

A conventional way to estimate weights in the literature is the inverse-probability weighting (IPW) method, where the weight of each sample is the corresponding inverse probability of receiving the treatment \citep{horvitz1952generalization,robins2000marginal,hirano2001estimation,hirano2003efficient}. It has been shown that this method can entirely remove the bias in the estimation of treatment effects when the true propensity score \citep{rosenbaum1983central}, defined as the conditional probability of receiving treatment given the covariates, is used. In order to apply the IPW method in practice, one needs to estimate the propensity score based on a presumed model, as the true propensity score is usually unknown in advance. However, as demonstrated by \cite{kang2007demystifying}, misspecification of the propensity score model can induce large biases in estimating treatment effects using IPW. This observation motivates several recent works to develop more robust ways to choose weights for the weighting methods, aiming to estimate the propensity score or the weight itself by directly comparing some prespecified moments/basis functions of covariates between the treatment groups  \citep{graham2012inverse,hainmueller2012entropy,imai2014covariate,zubizarreta2015stable,chan2016globally,zhao2016entropy,fan2016improving,wong2018kernel,zhao2019covariate,wang2020minimal}. These direct balancing weighting methods have been shown to work more robustly than the inverse-probability weighting method in practice. 

The good performance of the inverse-probability weighting or direct balancing weighting methods usually relies on strong regularity conditions for either the propensity score or the response functions. Specifically, the smoothness levels of the propensity score and the response functions must be at least larger than half of the covariate dimension in these methods \citep{chan2016globally,fan2016improving,wong2018kernel,wang2020minimal}. However, it is not immediately clear how to construct the weights in the weighting methods when the propensity score and the response functions are not as smooth as required and to what extent the treatment effects could be estimated by the weighting method in such non-smooth cases. Therefore, this paper aims to address these issues and develop a new framework of weighting methods to fill these needs.

We first investigate the weighting methods from a functional estimation perspective as a treatment effect is essentially a functional of the response and density functions. Through this perspective, we argue that the best way to estimate the ideal weights does not necessarily lead to the most efficient weighting estimator for treatment effects in the non-smooth case. In other words, the weights needed for covariate balancing could be different from the weights needed to estimate treatment effects when the response and density functions are non-smooth. A simple but useful example is given in Section~\ref{sc:warmup} to illustrate this point. Then, a natural question arises: how can we design a weighting method that directly targets the treatment effects estimation?

To answer this question, we introduce a new weighting framework called Weighting by a Uniform Transformer (WUNT). 
This new framework is motivated by an interesting observation about the uniform transformer, defined as a transformation mapping the covariate distribution in the control group to a uniform distribution. The uniform transformer gives us a straightforward and clean form of the weighting method, allowing us to directly make an accurate trade-off between the bias and variance for the treatment effects estimation. At first glance, such an observation may have little practical usage because one rarely knows the uniform transformer in advance, even if it exists. However, we show in this paper that it is possible to construct a data-driven uniform transformer from the covariates of control samples in a computationally efficient way. Furthermore, thanks to the data-driven uniform transformer, the weights in WUNT are customized for the treatment effects estimation, and the resulting weighting estimator is a simple kernel-based $U$-statistic. 

To demonstrate the merits of the newly proposed framework WUNT and the corresponding estimators, we study the theoretical properties under a nonparametric setting, especially when the response and density functions are non-smooth. Specifically, we show that the proposed estimator is consistent under very mild conditions. In addition, if the covariate density in the control group is known or can be estimated accurately, the minimax optimal converge rate for the mean square error of estimating the average treatment effect on the treated group is 
$$
n^{-{4(\alpha+\beta)\over d+2(\alpha+\beta)}}+n^{-1},
$$ 
where $n$ is the sample size, $d$ is the dimension of covariates, and $\alpha$ and $\beta$ are the smoothness levels of the response surfaces and density functions of the covariates, respectively. This result suggests that estimation of the treatment effects becomes more difficult when the response and density functions are less smooth. The converge rate presented here also appears in \cite{robins2008higher,robins2009semiparametric,robins2017minimax}. While \cite{robins2008higher,robins2017minimax} adopt higher-order influence functions to achieve an accurate trade-off between bias and variance, our result shows that the newly proposed weighting method is also able to do so because of the uniform transformer. To our best knowledge, this is the first minimax rate-optimal weighting method when the response and density functions are non-smooth. The practical merits of WUNT are further demonstrated through comprehensive simulation experiments. 

The rest of the paper is organized as follows. We first introduce the setting and give a brief review of weighting methods in Section~\ref{sc:weight}. Section~\ref{sc:new} presents the proposed framework of weighting, Weighting by a Uniform Transformer, and discusses several choices of the uniform transformer. In Section~\ref{sc:theo}, we investigate the theoretical properties of WUNT under a nonparametric setting. Finally, we conduct several numerical experiments in Section~\ref{sc:num} to demonstrate the practical performance of WUNT. All proofs and auxiliary results are relegated to the Supplement Material. 


\section{Problem Setting and Weighting Methods}
\label{sc:weight}

\subsection{Problem Setting and Notations}
\label{sc:model}

Suppose that the observed data $(\bX_i,Z_i,Y_i),\ i=1,\ldots,n$ are independent and identically distributed observations of $(\bX,Z,Y)$, where $\bX\in \RR^d$ are the observed covariates, $Z$ is a binary indicator variable for the treatment and $Y$ is the outcome of interest. Under the potential outcome framework for causal inference \citep{rubin1974estimating,imbens2015causal}, $Y^0$ and $Y^1$ are the potential outcomes when the individual is assigned to the treated ($Z=1$) or control group ($Z=0$). Then, the observed outcome can be written as $Y=(1-Z)Y^0+ZY^1$. Throughout this paper, we always assume the strong ignorability of the treatment assignment \citep{rosenbaum1983central}
\begin{equation}
\label{eq:ignor}
\{Y^0,Y^1\}\perp Z\ |\ \bX\qquad {\rm and}\qquad 0<\PP(Z=1|\bX)<1.
\end{equation}
Under this model, it is of interest to estimate the average treatment effect (ATE) or the average treatment effect on the treated group (ATT)
$$
\tau_{\rm ATE}=\EE(\mu_T(\bX)-\mu_C(\bX))\qquad {\rm and}\qquad \tau_{\rm ATT}=\EE(\mu_T(\bX)-\mu_C(\bX)|Z=1),
$$
where $\mu_T(\bX)$ and $\mu_C(\bX)$ are defined as
$$
\mu_T(\bX)=\EE(Y^1|\bX)\qquad {\rm and}\qquad \mu_C(\bX)=\EE(Y^0|\bX).
$$

For the sake of concreteness and illustration, we focus primarily on the average treatment effect on the treated group $\tau_{\rm ATT}$ in this paper. The techniques are also applicable to more generalized cases, e.g., Section~\ref{sc:ate} of the Supplement Material discusses robustly estimating the average treatment effect $\tau_{\rm ATE}$ under the new framework. If we adopt the following notation 
$$
f_T(\bX)=\PP(\bX|Z=1)\qquad {\rm and}\qquad f_C(\bX)=\PP(\bX|Z=0),
$$
then $\tau_{\rm ATT}$ can be written as 
$$
\tau_{\rm ATT}=\mu_{TT}-\mu_{CT}=\int \mu_T(\bX)f_T(\bX)d\bX - \int \mu_C(\bX)f_T(\bX)d\bX.
$$
It is natural to estimate the first term $\mu_{TT}$ by $\sum_{i=1}^nY_iZ_i/\sum_{i=1}^nZ_i$. The second term $\mu_{CT}$ is the major challenge in estimating $\tau_{\rm ATT}$ since the data with both response function $\mu_C(\bX)$ and sampling distribution $f_T(\bX)$ are inaccessible. Therefore, the main parameter of interest in this paper is $\mu_{CT}$.

\subsection{Weighting Methods}

A common method to estimate $\mu_{CT}$ is the covariates balancing or adjustment method.
This paper mainly focuses on the weighting methods \citep{rosenbaum1987model,hirano2003efficient}, which seek weights for each control sample so that covariates of weighted control samples are more similar to those of treated samples. Given the weights $w_i$ for each sample, $\mu_{CT}$ is estimated by the weighted mean
\begin{equation}
\label{eq:weightmethod}
\hat{\mu}_{CT}=\sum_{i=1}^nw_i(1-Z_i)Y_i.
\end{equation}
A standard choice of weights is based on the propensity score $w_i\propto \pi(\bX_i)/(1-\pi(\bX_i))$, where the propensity score is defined as $\pi(\bX)=\PP(Z=1|\bX)$ \citep{rosenbaum1983central}. Since the propensity score is usually unknown in advance, one needs to estimate the propensity score $\pi(\bX)$ by a parametric or nonparametric model. For example, a widely used parametric method for propensity score estimation is logistic regression. 

In order to make weighting methods more robust, several optimization-based weighting methods have been proposed recently \citep{imai2014covariate,zubizarreta2015stable,chan2016globally,fan2016improving,zhao2019covariate,wang2020minimal}. We call them direct balancing weighting methods in this paper. Instead of estimating the propensity score $\pi(\bX)$, direct balancing weighting methods aim to estimate weights $w_i\propto\pi(\bX_i)/(1-\pi(\bX_i))$ directly by comparing the moments/basis functions of covariates. 
More specifically, the weights $w_i$ of these methods can be calculated from the following optimization problem \citep{wang2020minimal}
\begin{equation}
\label{eq:dbw}
\begin{split}
\min_{w_i}\quad & \sum_{i=1}^n(1-Z_i)D(w_i)\\
{\rm s.t.}\quad & \left|\sum_{i=1}^nw_i(1-Z_i)\psi_l(\bX_i)-{1\over n_1}\sum_{i=1}^nZ_i\psi_l(\bX_i)\right|\le \Delta_l,\quad l=1,\ldots, L.
\end{split}
\end{equation}
Here, $D(\cdot)$ is a convex function of the weight, $n_0=\sum_{i=1}^n(1-Z_i)$, $n_1=\sum_{i=1}^nZ_i$, and $\psi_l(\bX), l=1,\ldots, L$ are some basis functions of the covariates and $\Delta_l\ge 0$ are the constraints for the imbalance in $\psi_l(\bX)$. The choices of basis functions play an important role in these methods \citep{fan2016improving,athey2018approximate,wang2020minimal}. In particular, the bias in the estimation of $\mu_{CT}$ can be well-adjusted when the response function belongs to the span of these basis functions approximately, i.e., there exists $a_1,\ldots,a_L$ such that
$$
\mu_C(\bX)\approx a_1\psi_1(\bX)+\ldots+a_L\psi_L(\bX).
$$

\subsection{A Functional Estimation Perspective}

In this section, we discuss the weighting methods from a functional estimation perspective as $\mu_{CT}=\int \mu_C(\bX)f_T(\bX)d\bX$ is essentially a bilinear functional of the response and density functions. The main intuition behind weighting is that the functional $\mu_{CT}$ can be rewritten as
\begin{equation}
	\label{eq:weightdecom}
	\mu_{CT}=\int \mu_C(\bX){f_T(\bX)\over f_C(\bX)}f_C(\bX)d\bX=\int\mu_C(\bX)f_C(\bX)w(\bX)d\bX,
\end{equation}
where the weighting function $w(\bX)$ is 
$$
w(\bX)={f_T(\bX)\over f_C(\bX)}={\pi(\bX)\over 1-\pi(\bX)}{\PP(Z=0)\over \PP(Z=1)}.
$$ 
Here, the weighting function at each $\bX_i$ can be treated as ideal weights for the weighting methods since $w(\bX)$ can make the distributions in the treated and control groups perfectly balanced. In practice, the weighting methods aim to estimate the weighting function at each $\bX_i$ by some estimator $\hat{w}(\bX_i)$ and then replace $w(\bX)$ by $\hat{w}(\bX)$ in \eqref{eq:weightdecom}. Inverse probability weighting and direct balancing weighting provide two different ways to estimate the ideal weights $w(\bX)$. Inverse probability weighting methods estimate the weighting function $w(\bX)$ through estimating the propensity score $\pi(\bX)$, while direct balancing weighting methods directly target the weighting function $w(\bX)$. Despite the difference in these methods, the common goal is to estimate the ideal weights $w(\bX)$.

On the other hand, the ultimate goal in treatment effects estimation is to estimate $\mu_{CT}$ rather than $w(\bX)$, so the weighting methods focusing on estimating the ideal weights can be seen as a plug-in estimator for $\mu_{CT}$ since $w(\bX)$ is replaced by $\hat{w}(\bX)$ in \eqref{eq:weightdecom}. However, such a plug-in strategy does not necessarily lead to an efficient estimator for the functional $\mu_{CT}$ \citep{lepski1999estimation,cai2011testing,robins2017minimax} because the best estimator for $w(\bX)$ might not be the most suitable for estimating $\mu_{CT}$. An explicit example is given to illustrate this point in Section~\ref{sc:warmup}. Therefore, a natural question arises: can we design the weights in \eqref{eq:weightmethod} that aim at estimating $\mu_{CT}$ directly? In this paper, we will see that this is possible; in fact, the weights needed for estimating $\mu_{CT}$ are over-debiased and variance-inflated estimators for $w(\bX)$ when the response function in the control is not smooth.

\subsection{A Warm-Up Example}
\label{sc:warmup}

To illustrate the idea, we start with a special case where the covariate $\bX$ is one-dimensional ($d=1$) and the distribution of $\bX$ in the control group $f_C(\bX)$ is the uniform distribution on $[0,1]$. In this case, the weighting function $w(\bX)$ becomes $f_T(\bX)$, so estimating the weights is equivalent to estimating the density of covariates in the treated group. To estimate the density $f_T$ without making any assumption on its parametric form, consider using one of the most commonly used density estimators -- the kernel density estimator
$$
\hat{f}_T(x)={1\over n_1h}\sum_{i=1}^n K\left(x-\bX_i \over h \right)Z_i={1\over n_1}\sum_{i=1}^n K_h\left(x-\bX_i  \right)Z_i,
$$
where $K(\cdot)$ is a one-dimensional kernel function and $h$ is the bandwidth. Standard analysis for kernel density estimators suggests that 
$$
\EE(\hat{f}_T(x)-f_T(x))^2\lesssim \underbrace{h^{2\beta}}_{Bias}+\underbrace{{1/(nh)}}_{Variance}
$$
if we assume $f_T(x)$ belongs to H{\"o}lder class $\Hcal^{\beta}([0,1])$; the formal definition of H{\"o}lder class is introduced in Section~\ref{sc:theo}. See \cite{tsybakov2008introduction} for a detailed proof. Therefore, if we aim to estimate the weighting function $w(\bX)$, the bandwidth $h$ should be chosen as $h\asymp n^{-1/(1+2\beta)}$, i.e., $h$ is of the same order as $n^{-1/(1+2\beta)}$. Is this choice of bandwidth also the most suitable one for estimating $\mu_{CT}$?

It seems reasonable to expect that the best way to estimate the weighting function $w(\bX)$ leads naturally to the best weighting estimator for $\mu_{CT}$ and hence the ATT. However, we now show that the bias and variance trade-off for estimating the weighting function $w(\bX)$ can be very different from estimating $\mu_{CT}$. If the weights in \eqref{eq:weightmethod} are replaced by the above kernel density estimator, the resulting weighting estimator for $\mu_{CT}$ is then
$$
\hat{\mu}_{CT}={\sum_{i_1,i_2=1}^nY_{i_1}(1-Z_{i_1})K_h\left(\bX_{i_1}-\bX_{i_2}  \right)Z_{i_2} \over \sum_{i_1,i_2=1}^n(1-Z_{i_1})K_h\left(\bX_{i_1}-\bX_{i_2}  \right)Z_{i_2}}.
$$
Since $\hat{\mu}_{CT}$ is a $U$-statistics, our analysis in Section~\ref{sc:theo} shows that 
$$
\EE(\hat{\mu}_{CT}-\mu_{CT})^2\lesssim  \underbrace{h^{2(\alpha+\beta)}}_{Bias}+\underbrace{1/n+{1/(n^2h)}}_{Variance},
$$
if we further assume the response surface in the control group $\mu_C(x)$ belongs to H{\"o}lder class $\Hcal^{\alpha}([0,1])$. The proof is omitted here since the result is a special case of Theorem~\ref{thm:upbd}. Unlike estimating the weighting function $w(\bX)$, the bias in estimating $\mu_{CT}$ also relies on the smoothness of the response function. The new bias and variance trade-off suggests that the optimal choice of the bandwidth $h$ for estimating $\mu_{CT}$ is 
$$
h\asymp \begin{cases}
	n^{-2/(1+2(\alpha+\beta))},&\qquad \alpha+\beta\le {1\over 2}\\
	[n^{-1},n^{-1/2(\alpha+\beta)}],&\qquad \alpha+\beta>{1\over 2}
\end{cases},
$$
where $h\asymp[a,b]$ means $h\asymp c$ for arbitrary $c\in [a,b]$. When the response function in the control group $\mu_C(x)$ is smooth enough, i.e., $\alpha\ge 1/2$, the optimal bandwidth for estimating the weighting function is also optimal for estimating $\mu_{CT}$. This explains why we can estimate the ATT efficiently through targeting the covariate balancing. On the other hand, if the response function $\mu_C(x)$ is non-smooth, i.e., $\alpha<1/2$, the optimal bandwidth for estimating $\mu_{CT}$ is much smaller than the optimal choice for estimating $w(\bX)$. In particular, we need to take the smoothness of the response function into account when we choose the optimal bandwidth for estimating $\mu_{CT}$. Putting differently, the optimal choice of $h$ for estimating $\mu_{CT}$ can result in a suboptimal estimator for $w(\bX)$, which is over-debiased and variance-inflated. This example not only illustrates the fact that the optimal estimator for $w(\bX)$ does not necessarily lead to an efficient estimator for $\mu_{CT}$ but also suggests a potential strategy to design the weights aiming for treatment effects estimation. 

\section{Weighting by a Uniform Transformer}
\label{sc:new}

\subsection{A Weighting Framework for Treatment Effects Estimation}

The warm-up example in the previous section suggests that the weights targeting the weighting function $w(\bX)$ may not lead to an efficient estimator for $\mu_{CT}$ and hence the ATT. So can we construct the weights that target $\mu_{CT}$ directly? In this section, we extend the idea in the warm-up example and introduce a new weighting framework that directly aims for treatment effects estimation. 

A unique feature of the warm-up example in Section~\ref{sc:warmup} is that the covariate follows a uniform distribution in the control group. Because of this property, the weights estimation problem is reduced to a density estimation problem. 
To apply this technique to covariates with any distribution, we can map the covariate distribution in the control group to a uniform distribution. Any such transformation $\Phi$ is referred to as a ``uniform transformer" in this paper. We assume the uniform transformer is known in the current section and leave the discussion on the construction of a uniform transformer to the next section. Let $\bU=\Phi(\bX)$ denote the data after transformation. Since $f^\Phi_C(\bU)$ is a uniform distribution, we can rewrite the propensity score-based weight $w_i$ as $$w_i\propto {\pi(\bX_i)\over 1-\pi(\bX_i)}\propto {f_T(\bX_i)\over f_C(\bX_i)}\propto {f^\Phi_T(\bU_i)\over f^\Phi_C(\bU_i)}\propto f^\Phi_T(\bU_i).$$ That is, to estimate the weights $w_i$, we only need to estimate the density $f^\Phi_T(\bU)$ at each $\bU_i$. However, as demonstrated in the warm-up example, the density estimation should be done carefully since the optimal tuning parameters targeting the density estimation itself can differ from the optimal choices for estimating the treatment effects. In other words, the key difference between our framework and the classical density estimation problems is the choice of tuning parameters. Under this new framework, the weights in \eqref{eq:weightmethod} can be constructed in two steps: (i) transform the covariates $\bX_i$ by a uniform transformer $\Phi$, (ii) estimate the weights by any density estimator with tuning parameters targeting the treatment effects estimation. The framework is summarized in Algorithm~\ref{ag:cool}, and we call it ``weighting by a uniform transformer" (WUNT).

\begin{algorithm}[h!]
	\caption{Weighting by Uniform Transformer (WUNT)}
	\label{ag:cool}
	\begin{algorithmic}
		\REQUIRE Data $\{(\bX_i,Z_i,Y_i)\}_{i=1}^n$.
		\ENSURE Weights $w_i$ and an estimator of $\mu_{CT}$.
		\STATE Construct the uniform transformer by $\{\bX_i\}_{i:Z_i=0}$ and apply transformation for all data $\bU_i=\Phi(\bX_i)$.
		\STATE Estimate $\hat{f}^\Phi_T(\bU)$ from $\{\bU_i\}_{i:Z_i=1}$.
		\STATE Evaluate the weights by $w_i={\hat{f}^\Phi_T(\bU_i) /\sum_{i:Z_i=0} \hat{f}^\Phi_T(\bU_i)}$ for $Z_i=0$.
		\STATE Assign the weights $w_i=1$ for $Z_i=1$.
		\STATE Estimate $\mu_{CT}$ by \eqref{eq:weightmethod}.
		\RETURN Weights $w_i$ and estimator $\hat{\mu}_{CT}$.
	\end{algorithmic}
\end{algorithm}

To conduct the density estimation in the second step, we introduce two of the most widely used nonparametric density estimators in the literature.
The first density estimator we consider here is the kernel density estimator, which has been widely used in many applications. The kernel density estimator is defined as 
$$
\hat{f}^\Phi_T(\bU)={1\over n}\sum_{i:Z_i=1}{1\over {\rm det}(H)}K\left(H^{-1}(\bU-\bU_i)\right)={1\over n}\sum_{i:Z_i=1}K_H\left(\bU-\bU_i\right),
$$
where $K(\cdot)$ is a kernel function and $H={\rm diag}(h_1,\ldots,h_d)$ is a  diagonal matrix of the bandwidths which controls the amount of smoothing. Let $K_H$ denote the scaled kernel with bandwidth matrix $H$ and write $K_H$ as $K_h$ when $h_1=\ldots=h_d=h$. In particular, we assume $K(\bX)=G(X_{(1)})\times \ldots\times G(X_{(d)})$, where $G(\cdot)$ is a univariate kernel $\int G(x)dx=1$. We call kernel $K$ an $\alpha$ order kernel if $\int x^tG(x)dx=0$ for any integer $t\le \alpha$ and $\int |x^\alpha G(x)|dx<\infty$. With the kernel density estimator, the final estimator of $\mu_{CT}$ in Algorithm~\ref{ag:cool} can be written as 
\begin{equation}
	\label{eq:kernel}
	\hat{\mu}_{CT}={\sum_{i_1,i_2=1}^nY_{i_1}(1-Z_{i_1})K_H(\Phi(\bX_{i_1})-\Phi(\bX_{i_2}))Z_{i_2}\over \sum_{i_1,i_2=1}^n(1-Z_{i_1})K_H(\Phi(\bX_{i_1})-\Phi(\bX_{i_2}))Z_{i_2}}.
\end{equation}

Another popular nonparametric density estimator is the projection density estimator. Given a series of orthonormal basis functions $\psi_l(\cdot)$, $l=1,\ldots,\infty$, $f^\Phi_T(\bU)$ can be decomposed as $f^\Phi_T(\bU)=\sum_{l=1}^{\infty} r_l\psi_l(\bU)$, where the coefficients are defined as  $r_l=\int f^\Phi_T(\bU)\psi_l(\bU)d\bU$. The projection method seeks to estimate $f^\Phi_T(\bU)$ with the first $L$ basis functions, i.e.,
$$
\hat{f}^\Phi_T(\bU)=\sum_{l=1}^{L} \hat{r}_l\psi_l(\bU),\qquad {\rm where}\ \  \hat{r}_l={1\over n_1}\sum_{i:Z_i=1}\psi_l(\bU_i).
$$
The projection density estimator then lead to the final estimator of $\mu_{CT}$
\begin{equation}
	\label{eq:projection}
	\hat{\mu}_{CT}={\sum_{i_1,i_2=1}^nY_{i_1}(1-Z_{i_1})K_L(\Phi(\bX_{i_1}),\Phi(\bX_{i_2}))Z_{i_2}\over \sum_{i_1,i_2=1}^n(1-Z_{i_1})K_L(\Phi(\bX_{i_1}),\Phi(\bX_{i_2}))Z_{i_2}},
\end{equation}
where $K_L(\bx,\by)=\sum_{l=1}^L\psi_l(\bx)\psi_l(\by)$ denotes a projection kernel defined by the orthonormal basis $\{\psi_l(\cdot):l=1,\ldots,L\}$ \citep{gine2016mathematical}. 

Although Algorithm~\ref{ag:cool} seems to suggest that $\hat{f}^\Phi_T$ is designed to estimate $f^\Phi_T$ at first glance, we would like to emphasize again that our ultimate goal is to estimate $\mu_{CT}$ instead of $f^\Phi_T$, so the choice of tuning parameters, $H$ in \eqref{eq:kernel} and $L$ in \eqref{eq:projection}, shall rely on the bias and variance trade-off in the final estimator $\hat{\mu}_{CT}$. The simple form of the estimators for $\mu_{CT}$ after applying the uniform transformer allows a more accurate trade-off between the bias and variance in estimating $\mu_{CT}$. We leave the detailed discussion of the tuning parameters $H$ and $L$ to Section~\ref{sc:theo}.

\subsection{Uniform Transformer}
\label{sc:unif}
\subsubsection{Rosenblatt's Uniform Transformer with Empirical Densities}
\label{sc:rosenblatt}


There are various possible options to transform a distribution into a uniform distribution. In this section, we focus on a uniform transformer proposed by \cite{rosenblatt1952remarks}. More concretely, given the density of covariates in the control group, $f_C(\bX)$, we consider the following transformation $\Phi:\Omega\to [0,1]^d$ 
\begin{equation}
	\label{eq:tran}
	\begin{split}
		\Phi(\bx)_{(1)}=&\PP_C(X_{(1)}\le x_{(1)}),\\
		\Phi(\bx)_{(2)}=&\PP_C(X_{(2)}\le x_{(2)}|X_{(1)}=x_{(1)}),\\
		\vdots&\\
		\Phi(\bx)_{(d)}=&\PP_C(X_{(d)}\le x_{(d)}|X_{(d-1)}=x_{(d-1)},\ldots,X_{(1)}=x_{(1)}),
	\end{split}
\end{equation}
where $\bx=(x_{(1)},\ldots,x_{(d)})\in \RR^d$ is a vector, $\bX=(X_{(1)},\ldots,X_{(d)})$ is a random vector with density $f_C(\bX)$ and the corresponding probability $\PP_C$. When $\bX$ is a continuous random vector, $\Phi(\bX)$ follows a uniform distribution on $[0,1]^d$. It is worth noting that the uniform transformer relies on the condition that $\bX$ is continuous. If some component of $\bX$ is discrete, a random perturbation can be added to the observed covariates in the preprocessing step, as discussed in \cite{brockwell2007universal}. In the following discussion, we assume $\bX$ is a continuous random vector; for more discussions on discrete random variables, see Section~\ref{sc:imple} in the Supplement Material. 

In practice, we usually do not have much knowledge about the density $f_C(\bX)$, so we have no access to the uniform transformer $\Phi$ defined in \eqref{eq:tran} and need to construct the uniform transformer from the data. One natural way to construct the uniform transformer is to first estimate the density of covariates in the control group $f_C(\bX)$ by some estimator $\tilde{f}_C(\bX)$ and then define the uniform transformer based on $\tilde{f}_C(\bX)$ following the transformation in \eqref{eq:tran}. 

Furthermore, the construction of uniform transformers becomes easier if the covariates have a special correlation structure. For instance, \eqref{eq:tran} suggests that the uniform transformer $\Phi$ only relies on each marginal distribution of $f_C(\bX)$ if components of $\bX$ are mutually independent. In other words, suppose $f_C(\bX)$ can be decomposed as $$f_C(\bX)=f_{C,1}(X_{(1)})\times \ldots\times f_{C,d}(X_{(d)}),$$ then it is sufficient to construct the uniform transformer by estimating each marginal distribution. 
When the uniform transformer is defined based on the marginal distributions, we call it a marginal uniform transformer. A similar idea can also be generalized to densities with a group-wise mutually independent structure.


In many applications, we could estimate the density function $f_C(\bX)$ or the corresponding marginal densities using a separated data set since there might be a much larger `unlabeled' data set of control samples besides the `labeled’ data set we observe. Here, the `unlabeled' control group refers to the samples for which we only observe the covariates $\bX_i$ and $Z_i=0$ but have no access to the corresponding response $Y_i$. For example, electronic health records (EHR) databases usually include an abundance of samples with automatically extracted covariates but without measuring the outcomes, as the measuring process is usually expensive and time-consuming \citep{gronsbell2017semi,chakrabortty2018efficient}. In such cases, one could use a large amount of `unlabeled' data to estimate density $f_C(\bX)$ accurately by any suitable density estimator and then construct the uniform transformer accordingly. After that, the uniform transformer can be applied to the `labeled’ data set to estimate the weights targeting treatment effects estimation.

\subsubsection{Adaptive Uniform Transformer}
\label{sc:aut}

Section~\ref{sc:rosenblatt} mainly focuses on Rosenblatt's uniform transformer and its empirical version with a density estimator. We provide a different angle to construct an empirical version of Rosenblatt's uniform transformer in this section. To illustrate the idea, we start with a one-dimensional case. When $d=1$, Rosenblatt's transformation is defined by the cumulative distribution function of $\bX$ in the control group. The cumulative distribution function can naturally be estimated by its empirical distribution, i.e.,
$$
\hat{\PP}_{C,n}(\bx)={1\over n_0}\sum_{i:Z_i=0} \bI(\bX_i\le \bx),
$$
where $\bI(\cdot)$ is an indicator function. If we plug $\hat{\PP}_{C,n}$ in Rosenblatt's transformation $\Phi$, the resulting transformation maps $\{\bX_i:Z_i=0\}$ to $\{1/n_0,\ldots, 1\}$. In other words, $\hat{\PP}_{C,n}$ help transform $\{\bX_i:Z_i=0\}$ to the grid points between 0 and 1. The benefit of this transformation is that it does not rely on estimating of $f_C(\bX)$, so it is computationally simple. 
Can we construct a uniform transformer in a similar fashion for a multi-dimensional case?

\begin{figure}[h!]
	\begin{center}
		\begin{tikzpicture}[scale=1.1]
			\draw[line width=2pt,black!80!white] (0,0) rectangle (4,4);
			
			\path[draw,line width=1.5pt,black!80!white,dashed] (0,1) -- (4,1);
			\path[draw,line width=1.5pt,black!80!white,dashed] (0,2.7) -- (4,2.7);
			
			\path[draw,line width=1.5pt,black!80!white,dashed] (1.1,0) -- (1.1,1);
			\path[draw,line width=1.5pt,black!80!white,dashed] (2.5,0) -- (2.5,1);
			\path[draw,line width=1.5pt,black!80!white,dashed] (1.4,1) -- (1.4,2.7);
			\path[draw,line width=1.5pt,black!80!white,dashed] (2.2,1) -- (2.2,2.7);
			\path[draw,line width=1.5pt,black!80!white,dashed] (0.9,2.7) -- (0.9,4);
			\path[draw,line width=1.5pt,black!80!white,dashed] (2.9,2.7) -- (2.9,4);
			
			\draw[fill=red!70!white] (0.5,0.2) circle (0.06);
			\draw[fill=red!70!white] (2.3,0.9) circle (0.06);
			\draw[fill=red!70!white] (3.7,0.6) circle (0.06);
			\draw[fill=red!70!white] (1.2,2.3) circle (0.06);
			\draw[fill=red!70!white] (2,1.6) circle (0.06);
			\draw[fill=red!70!white] (3.5,2.6) circle (0.06);
			\draw[fill=red!70!white] (0.7,3.3) circle (0.06);
			\draw[fill=red!70!white] (2.8,3.6) circle (0.06);
			\draw[fill=red!70!white] (3.9,3) circle (0.06);
			
			\draw[thick,black](-0.4,2)node{$X_{(1)}$};
			\draw[thick,black](2,-0.4)node{$X_{(2)}$};
		\end{tikzpicture}
	\end{center}
	\caption{An illustrative example for the construction of $\Phi$.}
	\label{fg:partition}
\end{figure}
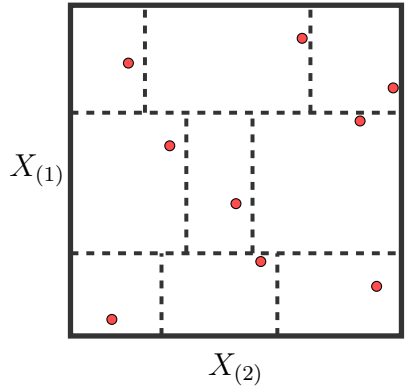

To apply this idea for a $d$-dimensional covariate $\bX$, we need to partition the data points evenly to $d$-dimensional grids. Assume the support of density is $\Omega=[0,1]^d$ and $n_0=N_0^d$ for some positive integer $N_0$ in this section. Based on $\{\bX_i\}_{i:Z_i=0}$, we define the following data-driven partition of $\Omega$
$$
\Omega=\bigcup_{j_1,\ldots,j_d=1}^{N_0} Q_{j_1,j_2,\ldots,j_d}=\bigcup_{j_1,\ldots,j_d=1}^{N_0} I_{j_1}\times I_{j_1,j_2}\times \ldots \times I_{j_1,j_2,\ldots,j_d}.
$$  
Here, each $Q_{j_1,j_2,\ldots,j_d}$ is a cube and each $I_{j_1,j_2,\ldots,j_k}$ for $k\le d$ is an interval. We construct the data-driven interval $I_{j_1,j_2,\ldots,j_k}$ in a hierarchical  way. We first construct $\{I_{j_1}\}_{ j_1=1}^{N_0}$, which is a partition of $[0,1]$ such that there are exactly $N_0^{d-1}$ points in each $I_{j_1}\times [0,1]^{d-1}$. After construction of $\{I_{j_1}\}_{ j_1=1}^{N_0}$, we are ready to construct $\{I_{j_1,j_2}\}_{ j_1,j_2=1}^{N_0}$. For each $j_1$, $\{I_{j_1,j_2}\}_{ j_2=1}^{N_0}$ is a  partition of $[0,1]$ such that there are exactly $N_0^{d-2}$ points in each $I_{j_1}\times I_{j_1,j_2}\times [0,1]^{d-2}$. The rest of $I_{j_1,j_2,\ldots,j_k}$ can be defined in a similar way. In doing so, each cube $Q_{j_1,j_2,\ldots,j_d}$ contains exactly one point. The idea is illustrated with an example of 9 data points on $[0,1]^2$ in Figure~\ref{fg:partition}.

Then, we can easily approximate the cumulative distribution function and conditional cumulative distribution functions based on the partition to form an empirical Rosenblatt's uniform transformer. With this construction, we can put exactly mass $1/n_0$ for each cube $Q_{j_1,j_2,\ldots,j_d}$. The formal result is summarized in the following proposition.

\begin{proposition}
	\label{prop:denunf}
	Let $\hat{\Phi}_D$ be the uniform transformer defined in \eqref{eq:tran} by replacing $f_C(\bX)$ with the following function	
	$$
	\ddot{f}(\bX)={1\over |Q_{j_1,\ldots,j_d}|n_0}
	S\left(X_{(1)}-M(I_{j_1})\over |I_{j_1}|\right)\times \ldots \times S\left(X_{(d)}-M(I_{j_1,\ldots,j_d})\over |I_{j_1,\ldots,j_d}|\right), \bX\in Q_{j_1,j_2,\ldots,j_d},
	$$
	where $\bX=(X_{(1)},\ldots, X_{(d)})$, $|\cdot|$ represent the volume of a cube or the length of an interval and $M(\cdot)$ is the middle point of an interval. Here, $S(\cdot)$ is a smooth kernel function defined on [-0.5,0.5] such that $S(-0.5)=S(0.5)=0$, $S(x)>0$ if $x\in(-0/5,0.5)$ and $\int_{-0.5}^{0.5} S(x)dx=1$. 
Then $\hat{\Phi}_D$ satisfies the following properties:
	\begin{enumerate}
		\item For each cube $Q_{j_1,j_2,\ldots,j_d}$, $1\le j_1,\ldots,j_d\le N_0$, 
		$$
		\hat{\Phi}_D(Q_{j_1,j_2,\ldots,j_d})=\left[{j_1-1\over N_0},{j_1\over N_0}\right)\times\ldots\times \left[{j_d-1\over N_0},{j_d\over N_0}\right).
		$$
		\item $\hat{\Phi}_D$ is a smooth map.
	\end{enumerate}
\end{proposition}
This proposition suggests that $\hat{\Phi}_D$ is able to map the covariates in the control group to an approximately uniform distribution on $[0,1]^d$, and we call it an adaptive uniform transformer. In particular, when $\Omega$ is on the real line ($d=1$), $\hat{\Phi}_D$ can been seen as a smoothed version of the empirical cumulative distribution function of $\{\bX_i\}_{i:Z_i=0}$. When there is no integer $N_0$ such that $n_0=N_0^d$, we can choose $N_0$ as the largest integer such that $N_0^d<n_0$ and follow a similar procedure as above to distribute the data points evenly in the grids. 

\section{Theoretical Properties}
\label{sc:theo}

We now turn to analyze the theoretical properties of our newly proposed framework. In this section, we consider uniform transformers introduced in Section~\ref{sc:unif} (either a uniform transformer constructed with a separate data set or an adaptive uniform transformer) and study the performance of both kernel and projection density estimators. In particular, our investigation focuses on H{\"o}lder class \citep{van1996weak}
$$
\Hcal^{\alpha}(\Omega)=\bigg\{f:\Omega\to \RR\big| \|f\|_{\alpha,\Hcal}\le M\bigg\},
$$
where the norm $\|f\|_{\alpha,\Hcal}$ is defined as
$$
\|f\|_{\alpha,\Hcal}=\max_{|k|\le \lfloor\alpha\rfloor}\sup_{\bx\in \Omega}|D^kf(\bx)|+\max_{|k|=\lfloor\alpha\rfloor}\sup_{\bx_1\ne\bx_2\in \Omega}{|D^kf(\bx_1)-D^kf(\bx_2)| \over \|\bx_1-\bx_2\|^{\alpha-\lfloor\alpha\rfloor}}.
$$
Here, $k=(k_1,\ldots,k_d)$ with $|k|=k_1+\ldots+k_d$ and the differential operator is defined as
$$
D^k={\partial^{|k|}\over \partial x_{(1)}^{k_1}\ldots  \partial x_{(d)}^{k_d}}.
$$
We assume the basis functions $\{\psi_l:l=1,\ldots,\infty\}$ in the projection density estimator form an orthonormal basis and satisfy 
\begin{equation}
\label{eq:basis}
|r_l|\le M_1l^{-(\alpha/d+1/2)}\qquad {\rm and}\qquad \sup_{\bx}|\psi_l(\bx)|\le M_2\sqrt{l},\qquad {\rm for \ any\ }l,
\end{equation}
where $M_1$ and $M_2$ are some constants and $r_l$ is the coefficient of some given function $g\in \Hcal^{\alpha}([0,1]^d)$, i.e., $g(\bx)=\sum_{l=1}^{\infty} r_l \psi_l(\bx)$. For example, the wavelet basis satisfies this property \citep{gine2016mathematical,liang2019estimating}. When $Z_i=0$, we write $Y_i=\mu_{C}(\bX_i)+\epsilon_i$, where $\EE(\epsilon_i|\bX_i)=0$. Through this section, we assume 
\begin{equation}
\label{eq:finitevar}
\EE(\epsilon_i^2):=\sigma(\bX_i)^2\le \sigma^2
\end{equation}
for some constant $\sigma^2$. 

We first investigate the performance of the proposed estimator when the uniform transformer in \eqref{eq:tran} is defined by some fixed density $\tilde{f}_C(\bX)$, which might be different from $f_C(\bX)$. The following theorem characterizes the convergence rate of the estimator. 

\begin{theorem}
	\label{thm:upbd}
	Let $\hat{\mu}_{CT}$ be the estimator defined in \eqref{eq:kernel} with an $\alpha+\beta$ order kernel or the one defined in \eqref{eq:projection} with basis function satisfying \eqref{eq:basis}. Suppose the uniform transformer is defined in \eqref{eq:tran} with some density $\tilde{f}_C(\bX)$. Assume  $\mu_C^\Phi\in \Hcal^\alpha([0,1]^d)$, $f_T^\Phi\in \Hcal^\beta([0,1]^d)$ and $f_C^\Phi\in \Hcal^\gamma([0,1]^d)$ with $0<\alpha,\beta<\gamma$. We further assume conditions \eqref{eq:ignor} and \eqref{eq:finitevar} hold. If we choose 
	 $h_1=\ldots=h_d=h=n^{-2/(d+2(\alpha+\beta))}$ in \eqref{eq:kernel} or $L=n^{2d/(d+2(\alpha+\beta))}$ in \eqref{eq:projection}, then there exists a constant $C_0$ such that
	$$
	\EE(\hat{\mu}_{CT}-\mu_{CT})^2 \le  C_0\left(n^{-{4(\alpha+\beta)\over d+2(\alpha+\beta)}}+n^{-1}+\Delta^2\right),
	$$
	where $\Delta$ is the difference between $\tilde{f}_C(\bX)$ and $f_C(\bX)$ in $L^2$ norm
	$$
	\Delta=\|\tilde{f}_C(\bX)-f_C(\bX)\|_2.
	$$
\end{theorem}

Theorem~\ref{thm:upbd} suggests that the performance of new estimators depends on the sum of smoothness levels of the response and density functions. Notably, they can still work well even when the response function is non-smooth ($\alpha < d/2$). In addition, it is worth noting that the optimal choice of tuning parameter $h$ or $L$ relies on the levels of smoothness of both the response and density functions. The optimal choice for estimating $f_T^\Phi$ ($h=n^{-1/(d+2\beta)}$ or $L=n^{d/(d+2\beta)}$) can lead to a suboptimal convergence rate in estimating $\mu_{CT}$. In other words, the best way to estimate $f^\Phi_T$ (hence the weights $w_i$) may not necessarily lead to the best estimator for $\mu_{CT}$. 

An immediate result of Theorem~\ref{thm:upbd} characterizes the performance of the proposed estimators when the density of covariates in the control group $f_C(\bX)$ is known. The formal result is summarized in the following corollary.
\begin{corollary}
	\label{cor:known}
	When the density of covariates in the control group $f_C(\bX)$ is known, the convergence rate of the estimator in \eqref{eq:kernel} or \eqref{eq:projection} is 
	\begin{equation}
		\label{eq:optirate}
		n^{-{4(\alpha+\beta)\over d+2(\alpha+\beta)}}+n^{-1}.
	\end{equation}
\end{corollary}

If the density of covariates in the control group $f_C(\bX)$ is not known, it can be estimated by some density estimator $\tilde{f}_C(\bX)$ with a separate data set, as discussed in Section~\ref{sc:rosenblatt}. In this case, the uniform transformer is constructed with this separate data set. The following corollary can further characterize the performance of new estimators. 

\begin{corollary}
	\label{cor:speupbd}
	Let $\tilde{f}_C(\bX)$ in Theorem~\ref{thm:upbd} be a density estimated by $N=cn^t$ control samples for some $t\ge 1$ and constant $c>0$. If $\|\tilde{f}_C(\bX)-f_C(\bX)\|_2\le N^{-\kappa/(d+2\kappa)}$ for some constant $\kappa$, we have the following results. When $\alpha+\beta\le d/2$ and $\kappa>2(\alpha+\beta) d/(td+(2t-4)(\alpha+\beta))$,
	then 
	$$
	\EE(\hat{\mu}_{CT}-\mu_{CT})^2 \le  Cn^{-{4(\alpha+\beta)\over d+2(\alpha+\beta)}}.
	$$
	When $\alpha+\beta>d/2$ and $\kappa>d/2(t-1)$,
	then
	$$
	{\sqrt{n}(\hat{\mu}_{CT}-\mu_{CT})\over \sqrt{V}}\to N(0,1),
	$$
	where $N(0,1)$ is standard normal distribution, $P=\PP(Z=1)$ and the variance $V$ is 
	$$
	V=\int {\mu_C^2(\bx)f_T(\bx)\over P}d\bx+\int {(\sigma^2(\bx)+\mu_C^2(\bx))f_T^2(\bx)\over (1-P)f_C(\bx)}d\bx-4\left(\int \mu_C(\bx)f_T(\bx)d\bx\right)^2.
	$$
\end{corollary}

We can conclude from this corollary that the converge rate in \eqref{eq:optirate} is still achievable as long as $f_C(\bX)$ can be estimated accurately. Now, we show that the converge rate in \eqref{eq:optirate} is actually sharp in terms of minimax optimality. More specifically, we assume $f_C(\bX)$ is known in advance and consider the following family of data distribution $(\bX,Z,Y)\sim F$ in $\Fcal_{\alpha,\beta}$ which is defined as 
$$
\Fcal_{\alpha,\beta}:=\bigg\{F: \mu_C^\Phi\in \Hcal^\alpha([0,1]^d),f_T^\Phi\in \Hcal^\beta([0,1]^d)\ {\rm and}\ \eqref{eq:ignor}, \eqref{eq:finitevar}\ {\rm hold}\bigg\}.
$$
Here, $\Phi$ is defined based on $f_C(\bX)$. 
\begin{theorem}
	\label{thm:lwbd}
	Consider estimating $\mu_{CT}$ on $\Fcal_{\alpha,\beta}$ with $\alpha,\beta>0$. Then there exists a constant $c_0$ such that 
	$$
	\inf_{\hat{\mu}_{CT}}\sup_{F\in \Fcal_{\alpha,\beta}}\EE(\hat{\mu}_{CT}-\mu_{CT})^2 \ge c_0\left(n^{-{4(\alpha+\beta)\over d+2(\alpha+\beta)}}+n^{-1}\right).
	$$
\end{theorem}
Theorems~\ref{thm:upbd} and \ref{thm:lwbd} together show that if  $f_C(\bX)$ is known, converge rate in \eqref{eq:optirate} is the minimax optimal rate of estimating $\mu_{CT}$.
This convergence rate is essentially the same as the one in \cite{robins2008higher,robins2009semiparametric,robins2017minimax}, which considers missing data models. When the response function and density function are not smooth, \cite{robins2008higher,robins2017minimax} adopt higher-order influence function methods to achieve this rate. Here, our results show that the weighting method is also able to achieve the minimax optimal rate.

In the above theorems, we assume $f_C(\bX)$ is known or can be estimated accurately by
separated data sets. We now show that the new estimator is still reliable without assumptions
on $f_C(\bX)$ if the uniform transformer is constructed as in Proposition~\ref{prop:denunf}.
\begin{theorem}
	\label{thm:consist}
	Let $\hat{\mu}_{CT}$ be the estimator defined in \eqref{eq:kernel} with an $\alpha+\beta$ order kernel or the one defined in \eqref{eq:projection} with basis function satisfying \eqref{eq:basis}. Assume the uniform transformer is defined in Proposition~\ref{prop:denunf}. Suppose  $\mu_C^\Phi\in \Hcal^\alpha([0,1]^d)$, $f_T^\Phi\in \Hcal^\beta([0,1]^d)$ with arbitrary $\alpha,\beta>0$ and conditions \eqref{eq:ignor}, \eqref{eq:finitevar} hold. For the kernel estimator, we choose bandwidth $h_1=\ldots=h_d=h$ satisfying $n^2h^d\to \infty$ and $h\to 0$. For the projection estimator, we choose the number of basis $L$ satisfying $n^2L^{-1}\to \infty$ and $L\to \infty$.
	Then,
	$$
	\hat{\mu}_{CT}\to_p \mu_{CT},\qquad {\rm as}\ n\to \infty.
	$$
\end{theorem}

Theorem~\ref{thm:consist} shows that this new uniform transformer can help build a consistent treatment effect estimator with no assumption on $f_C(\bX)$ and very mild conditions on $f_T^\Phi$ and $\mu_C^\Phi$. 
In this section, we focus mainly on the estimation of $\mu_{CT}$. All these results can naturally lead to the conclusion for the average treatment effect on the treated group, $\tau_{\rm ATT}$. Notably, when the density $f_C(\bX)$ is known, the minimax optimal rate of estimating $\tau_{\rm ATT}$ is 
$$
\inf_{\hat{\tau}_{\rm ATT}}\sup_{F\in \Fcal_{\alpha,\beta}}\EE(\hat{\tau}_{\rm ATT}-\tau_{\rm ATT})^2 \asymp n^{-{4(\alpha+\beta)\over d+2(\alpha+\beta)}}+n^{-1}.
$$
\section{Numerical Experiments}
\label{sc:num}

In this section, we study the numerical performance of our proposed framework by carrying out several simulation studies to estimate the average treatment effect on the treated group (ATT). 

\subsection{Comparison of Uniform Transformers}

In the first set of simulation studies, we compare four ways to construct the uniform transformer $\Phi$ from the control samples. More specifically, the adaptive uniform transformers are constructed in four different ways according to the following: if extra `unlabeled' control samples are used or not, and if the uniform transformers are based on joint or marginal distribution. To simulate the observed data, we draw $(W_1,...,W_5)\sim N((0.5,...,0.5),\Sigma)$ in the treated group and $(W_1,...,W_5)\sim N((0,...,0),\Sigma)$ in the control group, where $N$ represents the normal distribution and each entry of the covariance matrix $\Sigma$ is defined as $\Sigma_{ij}=\rho^{|i-j|}$. We vary $\rho$ from 0, 0.1, 0.2 and 0.3. The observed covariates of each sample is $\bX=(X_1,...,X_5)$ that $X_i=\exp(W_i)+W_i$. We consider two models for the outcome of interest: $Y_1=W_1^2W_2^2-2W_3^2W_4^2+\sum_{i=1}^5 W_i+\epsilon_1$ and  $Y_2=10\sum_{i=1}^3 W_i+100\prod_{i=1}^2\sin(2\pi W_i)+100\prod_{i=3}^5\cos(\pi W_i/2)+\epsilon_2$, where $\epsilon_i\sim N(0,1)$ follows independent standard normal distribution. We consider two density estimators -- the kernel density estimator and the projection density estimator, and denote their corresponding estimators of the ATT by $\hat{\tau}^K_{\rm ATT}$ and $\hat{\tau}^P_{\rm ATT}$, respectively. The sample size is $500$ for the treated group and $1000$ for the control group. The extra `unlabeled' control samples are drawn in the same way with sample size $10000$. The performances of different uniform transformers are evaluated by bias and root mean squared error (RMSE), calculated from $500$ replications of simulation experiments. 

The results are summarized in Table~\ref{tb:uniform}. These results show that when the covariate distribution of control samples is independent, the marginal uniform transformer works slightly better than the joint uniform transformer. This observation makes sense because when $f_C$ is an independent distribution, it is sufficient to construct the uniform transformer for each marginal distribution, as discussed in Section~\ref{sc:rosenblatt}. On the other hand, the joint uniform transformer is more robust when the observed covariates are correlated. In addition, the uniform transformer works in a better way when extra data is available. 

\begin{table}[h!]
	\centering
	\begin{tabular}{ccccccccccccccc}
		\hline\hline
		&	&	&&\multicolumn{5}{c}{No extra data} & & \multicolumn{5}{c}{With extra data}\\
		\cline{5-9}\cline{11-15}
		&	&	&&\multicolumn{2}{c}{Joint}& & \multicolumn{2}{c}{Marginal}& &\multicolumn{2}{c}{Joint}& &\multicolumn{2}{c}{Marginal}\\
		\cline{5-6}\cline{8-9}\cline{11-12}\cline{14-15}
		&	& $\rho$	&& Bias & RMSE && Bias & RMSE && Bias & RMSE && Bias & RMSE \\ 
		\hline
		\multirow{8}{*}{$Y_1$} & \multirow{4}{*}{$\hat{\tau}^K_{\rm ATT}$} &$0$&& -0.09 & 0.49 && -0.11 & 0.47 && 0.06 & 0.48 && -0.03 & 0.45 \\ 
		&&$0.1$& & -0.16 & 0.56 && -0.26 & 0.59 && 0.00 & 0.53 && -0.18 & 0.54 \\ 
		&&$0.2$ && -0.22 & 0.64 && -0.36 & 0.74 && -0.06 & 0.60 && -0.3 & 0.67\\ 
		&&$0.3$ &&-0.30 & 0.71 && -0.36 & 0.88 && -0.13 & 0.68 && -0.37 & 0.80 \\ 
		\cline{3-15}
		& \multirow{4}{*}{$\hat{\tau}^P_{\rm ATT}$} &$0$ && 0.18 & 0.56 && -0.06 & 0.51 && 0.03 & 0.53 && -0.06 & 0.52 \\ 
		&&$0.1$ && 0.13 & 0.61 && -0.11 & 0.58 && 0.00 & 0.58 && -0.11 & 0.58 \\ 
		&&$0.2$ && 0.11 & 0.68 && -0.14 & 0.66 && -0.01 & 0.66 && -0.14 & 0.66 \\ 
		&&$0.3$ && 0.09 & 0.75 && -0.16 & 0.75 && -0.02 & 0.74 && -0.16 & 0.75 \\
		\hline 
		\multirow{8}{*}{$Y_2$} & \multirow{4}{*}{$\hat{\tau}^K_{\rm ATT}$} &$0$ && 1.15 & 4.31 && 1.46 & 4.16 && 1.74 & 4.36 && 0.82 & 3.96 \\ 
		&&$0.1$ && 1.24 & 4.22 && 0.47 & 4.08 && 1.84 & 4.43 && 0.28 & 3.98 \\ 
		&&$0.2$& & 1.13 & 4.14 && -0.82 & 4.17 && 1.77 & 4.32 && -0.55 & 4.02 \\ 
		&&$0.3$ && 0.88 & 4.05 && -2.31 & 4.61 && 1.46 & 4.10 && -1.59 & 4.19\\ 
		\cline{3-15}
		& \multirow{4}{*}{$\hat{\tau}^P_{\rm ATT}$} &$0$ && 2.19 & 4.61 && 1.98 & 4.70 && 1.83 & 4.48 && 1.96 & 4.69 \\ 
		&&$0.1$ && 2.24 & 4.57 && 1.86 & 4.57 && 1.74 & 4.42 && 1.84 & 4.56\\ 
		&&$0.2$ && 2.23 & 4.57 && 1.71 & 4.44 && 1.65 & 4.31 && 1.69 & 4.43 \\ 
		&&$0.3$ && 2.13 & 4.43 && 1.51 & 4.19 && 1.55 & 4.16 && 1.49 & 4.18 \\ 
		\hline\hline
	\end{tabular}
	\caption{Comparison of uniform transformers under different covariance matrices.}
	\label{tb:uniform}
\end{table}

\subsection{Comparison of ATT Estimators}

The second set of simulation studies compares the newly proposed estimators with other existing methods under the above model. The four new estimators we consider here are: uniform transformer on joint distribution $+$ kernel density estimator, uniform transformer on marginal distribution $+$ kernel density estimator, uniform transformer on joint distribution $+$ projection density estimator, and uniform transformer on marginal distribution $+$ projection density estimator. We compare them with the inverse probability weighting estimator (IPW) with the propensity score estimated by random forests with \texttt{R} package \texttt{randomForest}, covariate balancing propensity score (CBPS) proposed by \cite{imai2014covariate} with \texttt{R} package \texttt{CBPS}, empirical balancing calibration weighting (CAL) by \cite{chan2016globally} with \texttt{R} package \texttt{ATE}, and stable weights (SBW) proposed by \cite{zubizarreta2015stable} with \texttt{R} package \texttt{sbw}. We still adopt bias and RMSE, calculated from 500 replications of simulation experiments again, as our measure of performance of these estimators for ATT. The data is generated in the same way as the first set of simulation studies with $\rho=0$. Results summarized in Table~\ref{tb:compestim1} suggest that our new proposed estimators perform better than the other methods in terms of bias and RMSE.

\begin{table}[h!]
	\centering
	\begin{tabular}{ccccccc}
		\hline\hline
		&&	\multicolumn{2}{c}{$Y_1$} &&\multicolumn{2}{c}{$Y_2$}\\
		\cline{3-4}\cline{6-7}
		&& Bias & RMSE && Bias & RMSE \\ 
		\hline
		Kernel+Joint && -0.10 & 0.51 && 0.94 & 4.38 \\ 
		Kernel+Marginal && -0.11 & 0.47 && 1.29 & 4.21 \\ 
		Projection+Joint && 0.20 & 0.58 && 2.01 & 4.57 \\ 
		Projection+Marginal && -0.04 & 0.53 && 1.85 & 4.65 \\ 
		IPW && 0.71 & 0.85 && 6.18 & 7.15 \\ 
		CBPS && 1.11 & 1.96 && 3.49 & 6.00 \\ 
		CAL && 1.07 & 1.64 && 3.93 & 5.99 \\ 
		SBW && 0.45 & 0.84 && 3.4 & 5.41 \\ 
		\hline\hline
	\end{tabular}
	\caption{Comparison of different ATT estimators on model $Y_1$ and $Y_2$.}
	\label{tb:compestim1}
\end{table}

\subsection{Comparison of ATT Estimators with Different Sample Sizes}

In the third set of simulation studies, we further compare the eight estimators of ATT in the second set of simulation studies and assess their performance with different sample sizes. In this set of simulation experiments, the data is simulated based on the example in \cite{kang2007demystifying}. More concretely, we draw $W = (W_1,W_2,W_3,W_4)$ from $N((0,0,0,0),I)$, where $I$ is a $4\times4$ identity matrix and consider the following two models of the outcome of interest: $Y_3=210+27.4W_1+13.7W_2+13.7W_3+13.7W_4+\epsilon_3$ (the same with \cite{kang2007demystifying}) and $Y_4=(4W_1+2W_2)/(\exp(W_3)+4\sqrt{|W_4|})+2W_3+W_4+\epsilon_4$. Here, $\epsilon_i$ also follows independent standard normal distribution. Each sample is assigned to the treated group with probability (i.e., true propensity score) $1/(1+\exp(W_1-0.5W_2+0.252_3+0.1W_4))$. Instead of observing the covariates $W$, we are able to observe only the transformed data $X_1 = \exp(W_1/2)$, $X_2 = W_2/(1 + \exp(W_1))+10$, $X_3 =(W_1W_3/25 + 0.6)^3$ and $X_4 = (W_2 + W_4 + 20)^2$. In order to assess the effect of sample size, we vary it from $1000$, $2000$, and $5000$. Similar to the previous two simulation studies, bias and RMSE based on 500 replications of simulation experiments are summarized in Table~\ref{tb:compestim2}. The results show that our new estimators generally outperform other existing methods. The only exception is the kernel density estimator equipped with a uniform transformer based on marginal distribution. The reason is that the observed covariates are highly dependent and the kernel density estimator seems to be sensitive to the correlation among covariates. Table~\ref{tb:compestim2} also suggests that the new estimators based on the kernel density estimator can constantly reduce the bias as the sample size increases. 

\begin{table}[h!]
	\centering
	\begin{tabular}{cccccccccc}
		\hline\hline
		\multirow{2}{*}{$Y_3$}	&	&\multicolumn{2}{c}{$n=1000$}&&\multicolumn{2}{c}{$n=2000$}&&\multicolumn{2}{c}{$n=5000$}\\\cline{3-4}\cline{6-7}\cline{9-10}
		&& Bias & RMSE && Bias & RMSE && Bias & RMSE \\ 
		\hline
		Kernel+Joint && -7.90 & 8.21 && -6.32 & 6.51 && -4.50 & 4.62 \\ 
		Kernel+Marginal && -9.88 & 10.16 && -9.46 & 9.62 && -8.88 & 8.96 \\ 
		Projection+Joint && -4.26 & 4.48 && -4.39 & 4.49 && -4.27 & 4.30 \\ 
		Projection+Marginal && -4.00 & 4.15 && -4.01 & 4.09 && -3.98 & 4.00 \\ 
		IPW && -10.13 & 10.26 && -9.96 & 10.02 && -9.61 & 9.64 \\ 
		CBPS && -5.35 & 5.56 && -5.40 & 5.51 && -5.31 & 5.37 \\ 
		CAL && -4.36 & 4.49 && -4.43 & 4.49 && -4.37 & 4.40 \\ 
		SBW && -7.22 & 7.32 && -7.40 & 7.44 && -7.4 & 7.42 \\ 
		\hline
		\hline
		\multirow{2}{*}{$Y_4$}	&	&\multicolumn{2}{c}{$n=1000$}&&\multicolumn{2}{c}{$n=2000$}&&\multicolumn{2}{c}{$n=5000$}\\\cline{3-4}\cline{6-7}\cline{9-10}
		&& Bias & RMSE && Bias & RMSE && Bias & RMSE \\ 
		\hline
		  Kernel+Joint && -0.40 & 0.44 && -0.31 & 0.34 && -0.22 & 0.24 \\ 
		Kernel+Marginal && -0.67 & 0.71 && -0.66 & 0.68 && -0.65 & 0.66 \\ 
		Projection+Joint && -0.38 & 0.41 && -0.38 & 0.39 && -0.37 & 0.38 \\ 
		Projection+Marginal && -0.38 & 0.4 && -0.38 & 0.39 && -0.39 & 0.39 \\ 
		IPW && -0.63 & 0.64 && -0.61 & 0.62 && -0.58 & 0.59 \\ 
		CBPS && -0.56 & 0.59 && -0.57 & 0.59 && -0.57 & 0.58 \\ 
		CAL && -0.46 & 0.49 && -0.47 & 0.49 && -0.48 & 0.48 \\ 
		SBW && -0.64 & 0.66 && -0.66 & 0.67 && -0.66 & 0.67 \\ 
		\hline\hline
	\end{tabular}
	\caption{Comparison of different ATT estimators on model $Y_3$ and $Y_4$.}
	\label{tb:compestim2}
\end{table}

\subsection{Comparison of Computation Complexity}

We compare the computation time of these eight methods in the last set of simulation studies. In particular, we consider two optimization solvers for SBW: quadpros (the default choice) and mosek (a commercial solver available from \url{https://www.mosek.com/}). We record the average time from 10 replications of the model $Y_3$ with sample sizes $1000$, $2000$, $5000$, and $10000$ and summarize them in Table~\ref{tb:time}. All these algorithms are evaluated with the same laptop (Intel Core i5 @2.3 GHz/8GB). From Table~\ref{tb:time}, we can conclude that the new estimators based on the projection density estimator can be computed efficiently. The main computation obstacle of the new estimators based on the kernel density estimator is the kernel $U$ statistics which have $O(n^2)$ computation complexity. It is also interesting to note that the uniform transformer can be constructed in a very short time ($<0.66$s even when $n=10000$) and thus can be applied on a larger scale dataset. 

\begin{table}[h!]
	\centering
	\begin{tabular}{cccccc}
		\hline\hline
		&& $n=1000$ & $n=2000$ & $n=5000$ & $n=10000$ \\ 
		\hline
		Kernel+Joint && 0.40 & 1.36 & 7.98 & 31.08 \\ 
		Kernel+Marginal && 0.36 & 1.31 & 8.01 & 32.19 \\ 
		Projection+Joint && 0.06 & 0.12 & 0.30 & 0.66 \\ 
		Projection+Marginal && 0.04 & 0.11 & 0.49 & 1.85 \\ 
		IPW && 0.33 & 0.73 & 2.24 & 4.61 \\ 
		CBPS && 0.31 & 0.64 & 1.67 & 4.23 \\ 
		CAL && 0.14 & 0.27 & 0.67 & 1.41 \\ 
		SBW(quadpros) && 0.14 & 1.19 & 20.60 & 152.18 \\ 
		SBW(mosek) && 0.03 & 0.04 & 0.08 & 0.19 \\ 
		\hline\hline
	\end{tabular}
	\caption{Comparison of different ATT estimators in terms of the computation time, which is shown in seconds.}
	\label{tb:time}
\end{table}

\section{Concluding Remarks}
\label{sc:dis}

In this paper, we propose a novel framework of weighting methods, weighting by a uniform transformer (WUNT), for treatment effects estimation. Unlike the existing weighting methods, the new framework utilizes a data-driven uniform transformer to the observed covariates, which transforms the covariate distribution in the control group to a uniform distribution. In doing so, we design the weights in WUNT to be particularly suitable for treatment effects estimation, and the final estimator is a simple kernel-based $U$-statistic. We also study the theoretical properties of the newly proposed framework under a nonparametric setting. Our investigation shows that, with weights chosen by WUNT, the weighting method is able to achieve the minimax optimal rate of estimating the average treatment effect on the treated group even under low regularity conditions. In particular, the tuning parameter in WUNT needs to be chosen based on the smoothness levels of both the response and density functions to achieve an accurate trade-off between bias and variance under low regularity conditions. Some implementation suggestions are given in Section~\ref{sc:implesugg} of the Supplement Material.

Although the main focus of this paper is the estimation of the average treatment effect on the treated group, the techniques are readily applicable to more generalized cases. For example, the new framework WUNT can also robustly estimate the average treatment effect; see Section~\ref{sc:ate} of the Supplement Material for more details. In addition, most of this paper assumes the observed covariates are continuous random variables, while there are some discrete covariates in many applications. WUNT, as a general framework, can be easily generalized to discrete random variables if we consider an extra preprocessing step for discrete random variables. See detailed discussion in Section~\ref{sc:imple} of the Supplement Material.

\bibliographystyle{plainnat}
\bibliography{WUNT}

\end{document}